\documentclass{article}
\usepackage{spconf,amsmath,graphicx}

\usepackage{enumitem}
\setlist{nosep, leftmargin=14pt}

\usepackage{booktabs}
\usepackage{amssymb}
\usepackage{makecell}
\usepackage{bm}
\usepackage{xcolor}
\usepackage{nicematrix}

\definecolor{Gray}{gray}{0.85}

\title{Improved Multimodal Fusion for Small Datasets with\linebreak Auxiliary Supervision}
\name{Gregory Holste$^{1,2}$, Douwe van der Wal$^2$, Hans Pinckaers$^2$, Rikiya Yamashita$^2$}

\address{\textit{Akinori Mitani$^2$, Andre Esteva$^2$}\\ \\ $^1$ The University of Texas at Austin, Austin, TX USA\\ $^2$ Artera, Inc, Mountain View, CA USA}

\begin{document}
%
\maketitle
\begin{abstract}

Prostate cancer is one of the leading causes of cancer-related death in men worldwide. Like many cancers, diagnosis involves expert integration of heterogeneous patient information such as imaging, clinical risk factors, and more. For this reason, there have been many recent efforts toward deep multimodal fusion of image and non-image data for clinical decision tasks. Many of these studies propose methods to fuse learned features from each patient modality, providing significant downstream improvements with techniques like cross-modal attention gating, Kronecker product fusion, orthogonality regularization, and more. While these enhanced fusion operations can improve upon feature concatenation, they often come with an extremely high learning capacity, meaning they are likely to overfit when applied even to small or low-dimensional datasets. Rather than designing a highly expressive fusion operation, we propose three simple methods for improved multimodal fusion with small datasets that \textit{aid optimization by generating auxiliary sources of supervision during training}: \textbf{extra supervision}, \textbf{clinical prediction}, and \textbf{dense fusion}. We validate the proposed approaches on prostate cancer diagnosis from paired histopathology imaging and tabular clinical features. The proposed methods are straightforward to implement and can be applied to any classification task with paired image and non-image data.

\end{abstract}
\section{Introduction}
\label{sec:intro}

\begin{figure*}[!ht]
    \centering
    \includegraphics[scale=0.75]{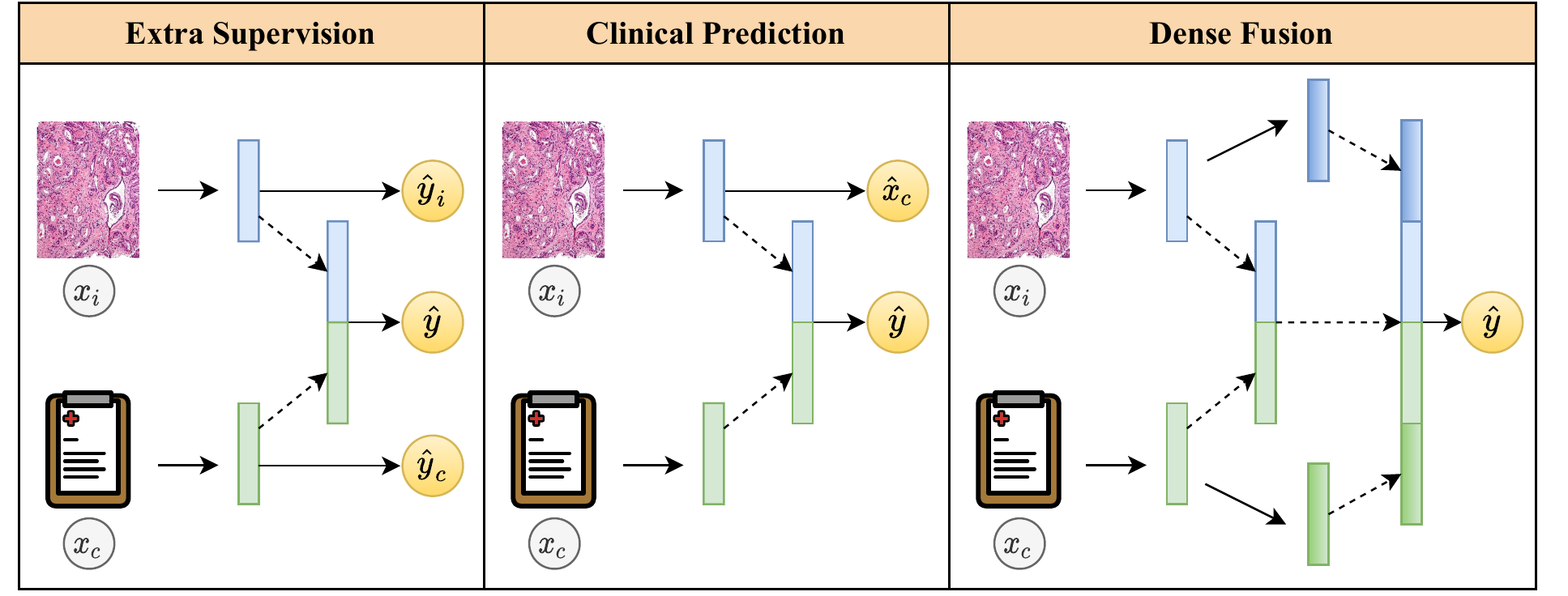}
    \caption{Diagram of proposed multimodal fusion approaches for small or low-dimensional datasets. \textbf{Extra supervision} (left) generates additional modality-specific predictions, \textbf{clinical prediction} (middle) uses the learned image representation to directly predict the associated non-image inputs, and \textbf{dense fusion} (right) encourages dense interaction of features by joining information from different stages of the network.}
    \label{fig:architectures}
\end{figure*}

Prostate cancer is the third most frequently diagnosed cancer worldwide and one of the leading causes of cancer-related death in men \cite{bray2018global,ward2019annual}. Diagnosis and risk stratification is primarily performed by manual analysis of histopathology imaging from tissue biopsy in the context of patient history, and risk factors like age and prostate-specific antigen (PSA) level \cite{schaeffer2021nccn}. Like many cancers, diagnosis relies on the integration of multiple streams of heterogeneous information about the patient. Given this clinical reality and the ongoing success of deep learning techniques for automated diagnosis, there has been recent attention toward multimodal fusion for automated cancer risk stratification; this involves using deep learning techniques to jointly learn from patient imaging (radiological, histopathological, or other), patient risk factors, and other streams of data like lab results and genomic profiles \cite{mobadersany2018predicting,cheerla2019deep,Holste_2021_ICCV}. In this line of research, a central open question remains: what is the best way to fuse information from each modality?

In their review of multimodal fusion approaches for imaging and electronic health record (EHR) data, Huang \textit{et al.}\ \cite{huang2020fusion} divide approaches into three categories: early fusion (joining inputs from each modality), joint fusion (joining learned intermediate features from each modality), and late fusion (joining predicted probabilities from each modality). Most studies find that some form of joint fusion (often called ``\textbf{late joint fusion}"), combining learned modality-specific feature vectors to generate a single multimodal representation, to be most effective \cite{huang2020fusion,cui2022deep}. Given, for example, an image representation and a clinical data representation, there are many ways to aggregate these two feature vectors -- the most common being concatenation and elementwise averaging. Observing room for improvement, researchers have proposed more complex and expressive fusion operations \cite{chen2020pathomic,braman2021deep} and architectures \cite{chen2021multimodal,huang2021gloria} to optimally join multimodal representations. For example, Chen \textit{et al.}\ \cite{chen2020pathomic,chen2021multimodal} use cross-modal attention gating and Kronecker product fusion to aggregate genomic and histopathology imaging features for cancer outcome prediction. Braman \textit{et al.}\ \cite{braman2021deep} build upon this by adding an orthogonality loss that encourages unimodal representations to provide complementary information during fusion.

While these fusion methods have shown significant improvements on diagnosis tasks, the resulting models often have a very high learning capacity (large number of learnable parameters). Thus when applied to small datasets or low-dimensional inputs, these methods are more likely to overfit to spurious patterns in the training data \cite{hastie2009elements}. We hypothesize that another avenue toward improved generalization \textit{without dramatically increasing model capacity} is to ease model optimization with multiple sources of supervision. In this work, we propose three straightforward methods that aid optimization by generating auxiliary sources of supervision: extra supervision, clinical prediction, and dense fusion. We validate these approaches on prostate cancer classification using paired histopathology image features and tabular clinical risk factors. Our experiments show that a combination of these three methods improves upon late joint fusion (with concatenation \textit{or} Kronecker product fusion) and can be readily applied to any task with paired image and non-image data.

\section{Materials \& Methods}
\label{sec:methods}

\subsection{Dataset Description}

This study uses data collected from five prospective, randomized phase III clinical trials of men with localized prostate cancer conducted by NRG Oncology: NRG/RTOG-9202, 9408, 9413, 9910, and 0126
\cite{jones2011radiotherapy,michalski2018effect,pisansky2015duration,horwitz2008ten,lawton2007update}; a subset of $4,581$ patients was made available for machine learning research in this work. For each patient, we use paired histopathology imaging and clinical risk factor data to predict distant metastasis (DM), the binary event that cancer spreads from the original tumor site. Image features were pre-extracted by a self-supervised learning model \cite{esteva2022prostate} so that each slide is represented by a lower-dimensional ``bag" of image features $\bm{\bm{x_i}} \in \mathbb{R}^{K \times 128}$, where $K$ is the number of fixed-size (256 \texttimes \ 256) patches. Each associated clinical input vector $\bm{x_c} \in \mathbb{R}^6$ includes six tabular descriptors of the patient: age, PSA, T-stage, and pathologist-determined Gleason scores and patterns \cite{gleason1974prediction}.
In total, DM occurs in 12.2\% of the 4,581 patients. For full details on data acquisition and image feature extraction, please refer to Esteva \textit{et al.}\ \cite{esteva2022prostate}.

\subsection{Fusion Methods for Small Datasets}
\label{sec:fusion_methods}

The standard ``late joint fusion" method of multimodal classification involves learning modality-specific representations, fusing those feature vectors into a single multimodal representation, then using this feature vector to perform classification. Concretely, let $\bm{x_i}$ represent a patient's histopathology imaging and $\bm{x_c}$ represent the patient's associated clinical information. Given image encoder $f_i(\cdot)$ and clinical encoder $f_c(\cdot)$, we compute representations $\bm{h_i} = f_i(\bm{x_i})$ and $\bm{h_c} = f_c(\bm{x_c})$, where $i$ refers to the imaging modality and $c$ refers to the clinical (or non-imaging) modality. Then image and non-image representations are fused via concatenation: $\bm{h} = \textrm{concat}([\bm{h_i}, \bm{h_c}])$. Finally, a classifier $g(\cdot)$ is used to generate a final prediction of the outcome via $\hat{y} = g(\bm{h})$. This model is optimized by minimizing the loss $\mathcal{L} = \ell(y, \hat{y})$, where $\ell(\cdot)$ is the binary cross-entropy loss function.

Building on this baseline, we present three methods that aid optimization by adding auxiliary sources of supervision during training: \textbf{extra supervision}, \textbf{clinical prediction}, and \textbf{dense fusion} (Figure \ref{fig:architectures}).

\subsubsection{Extra Supervision}
\label{sec:extra_supervision}

While the baseline late joint fusion approach generates a single prediction based on fused image and non-image features, it does not directly encourage the modality-specific representations to be predictive of the outcome. We can remedy this by adding classification heads on top of each modality-specific representation. Given image-only classifier $g_i(\cdot)$ and clinical-only classifier $g_c(\cdot)$, we additionally compute $\hat{y}_i = g_i(\bm{h_i})$ from image features alone and $\hat{y}_c = g_c(\bm{h_c})$ from clinical features alone. This approach, \textbf{extra supervision}, now generates three predictions of the outcome, which can together be used to optimize the entire network via the loss $\mathcal{L} = \ell(y, \hat{y}) + \ell(y, \hat{y}_i) + \ell(y, \hat{y}_c)$. This way, additional supervisory signal from the modality-specific features can flow back to optimize all parameters in the network. Similar approaches have appeared in Kawahara \textit{et al.}\ \cite{kawahara2018seven} and Holste \textit{et al.}\ \cite{Holste_2021_ICCV}.

\subsubsection{Clinical Prediction}
\label{sec:clinical_prediction}

In \textbf{clinical prediction}, we build upon the standard late jointfusion approach by using the learned image representation to directly predict (regress) the associated non-image input features. Specifically, we compute $\bm{\hat{x}_c} = g_{i \rightarrow c}(\bm{h_i})$ with an additional classification head $g_{i \rightarrow c}(\cdot)$. In addition to the predicted outcome, this new predicted clinical feature vector $\bm{\hat{x}_c}$ serves as an auxiliary source of supervision during training via the loss $\mathcal{L} = \ell(y, \hat{y}) + \textrm{sim}(\bm{x_c}, \bm{\hat{x}_c})$, where $\ell(\cdot)$ is the binary cross-entropy loss and $\textrm{sim}(\cdot)$ is the mean squared error (MSE) loss for continuous inputs. This approach can be interpreted as an auxiliary prediction task that aligns the learned image representations with the clinical non-image features. Since the image and non-image ``views" of the patient are expected to be correlated, this alignment of modalities may enable more optimal fusion of image and non-image information.

\subsubsection{Dense Fusion}
\label{sec:dense_fusion}

Inspired by Hu \textit{et al.}\ \cite{hu2019dense}, \textbf{dense fusion} improves upon standard late joint fusion by allowing for denser interaction between modalities during training. As in late joint fusion, we learn image-only and non-image-only representations via $\bm{h_i^{(1)}} = f_i^{(1)}(\bm{x_i})$ and $\bm{h_c^{(1)}} = f_c^{(1)}(\bm{x_c})$, and then those representations are fused to form $\bm{h^{(1)}} = \textrm{concat}([\bm{h_i^{(1)}}, \bm{h_c^{(1)}}])$; the superscript $(1)$ is denotes that this is the first intermediate layer of the network. However, instead of placing a classifier on top of the fused representation $\bm{h^{(1)}}$, we instead learn a deeper representation of the image and non-image features with $\bm{h_i^{(2)}} = f_i^{(2)}(\bm{x_i})$ and $\bm{h_c^{(2)}} = f_c^{(2)}(\bm{x_c})$, where $f_i^{(2)}$ and $f_c^{(2)}$ are fully-connected layers. We then form a final fused representation that not only aggregates the image-only and clinical-only features (as in late joint fusion), but also incorporates the fused representation from the \textit{previous} stage of the network: $\bm{h^{(2)}} = \textrm{concat}([\bm{h_i^{(2)}}, \bm{h_c^{(2)}}, \bm{h^{(1)}}]).$ Finally, a classifier $g(\cdot)$ is used to generate $\hat{y} = g(\bm{h^{(2)}})$, and the model is trained by optimizing the same binary cross-entropy loss as in late joint fusion. This allows for dense interaction of features from each modality, aggregating information across different stages of the network.

\begin{figure}[!ht]
    \centering
    \includegraphics[scale=0.75]{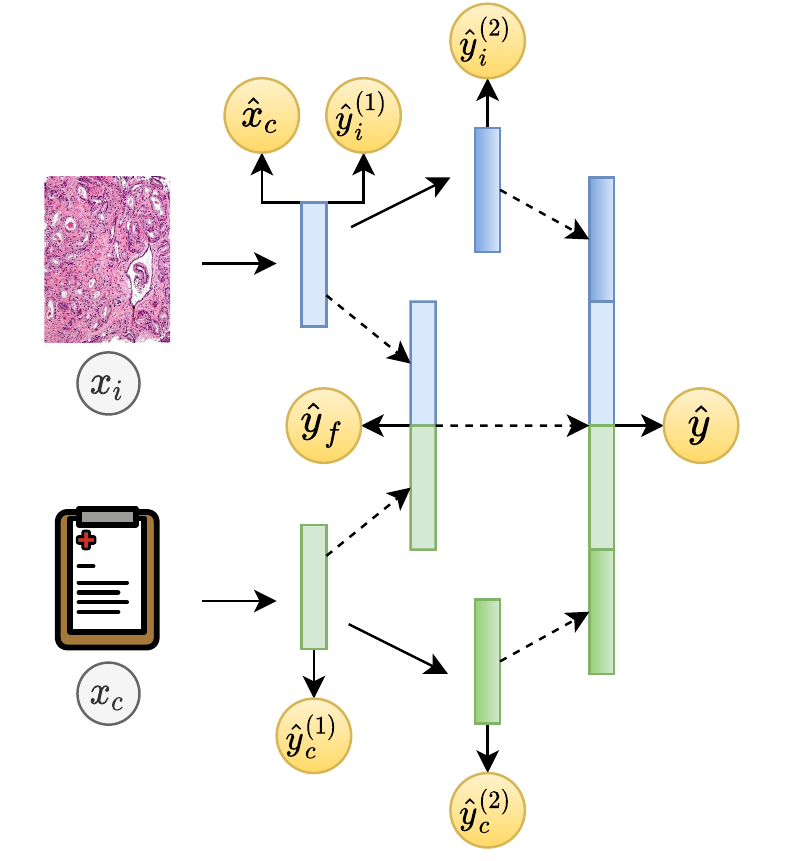}
    \caption{Multimodal fusion architecture with all three approaches combined: extra supervision, clinical prediction, and dense fusion. The model is trained by optimizing the sum of six cross-entropy losses (one for each predicted $\hat{y}$) and one mean squared error loss (for regressing clinical features $\bm{\hat{x}_c}$).}
    \label{fig:all}
\end{figure}

While dense fusion does not explicitly generate additional sources of supervision like the other two approaches, it can be combined with extra supervision and clinical prediction to obtain even more sources of supervision than either approach alone. Dense fusion provides multiple intermediate feature vectors; when combined with extra supervision, these feature vectors are used to generate \textit{additional} outcome predictions. Further, \textbf{all three methods are complementary and can be applied in any combination}. For example, Figure \ref{fig:all} depicts an architecture combining all three proposed approaches; this model minimizes the loss $\mathcal{L} = \ell(y, \hat{y}) + \ell(y, \hat{y}_f) + \ell(y, \hat{y}_i^{(1)}) + \ell(y, \hat{y}_i^{(2)}) + \ell(y, \hat{y}_c^{(1)}) + \ell(y, \hat{y}_c^{(2)}) + \textrm{sim}(\bm{x_c}, \bm{\hat{x}_c})$.
For all models, the final prediction at inference time is taken to be $\hat{y}$, the output learned from the final fused multimodal representation.



\subsection{Implementation Details}

As in Esteva \textit{et al.}\ \cite{esteva2022prostate}, all six clinical features were treated as numeric and standardized. To generate a 128-dimensional representation for each bag of imaging features $\bm{x_i} \in \mathbb{R}^{K \times 128}$, attention weights were learned to modulate certain features via elementwise multiplication, then sum pooling was utilized across the patch dimension. All models were trained with the AdamW optimizer \cite{loshchilov2019} with learning rate 0.0075, batch size 128, and dropout 0.9 on classification heads. Each model was trained for 100 epochs with class-balanced resampling per fold to combat imbalance. For the first 20 epochs, only the clinical encoder $f_c(\cdot)$ and classifier $g_c(\cdot)$ were optimized; then, the parameters of $f_c(\cdot)$ were frozen while all other parameters were optimized for the remaining 80 epochs.

\begin{table}[!ht]
    \centering
    \renewcommand{\arraystretch}{1.1}




    \begin{NiceTabular}{@{}cccc@{}}[colortbl-like]
    \toprule
    \makecell{\textbf{Fusion}\\\textbf{Operation}} & \makecell{\textbf{Auxiliary}\\\textbf{Supervision}} & \textbf{\# Params} & \textbf{AUC} \\ \midrule
    \rowcolor{Gray} \ \ Concatenation & & 17.1K & 0.781 $\pm$ 0.024\ \  \\
    \ \ Kronecker & & 66.3K & 0.770 $\pm$ 0.018\ \  \\
    \ \ Concatenation & \checkmark & 43.1K & \textbf{0.792 $\pm$ 0.014}\ \  \\
    \ \ Kronecker & \checkmark & 207.1K & 0.781 $\pm$ 0.013\ \  \\ \bottomrule
    \end{NiceTabular}

    \caption{Distant metastasis (DM) classification results with different fusion techniques and our proposed auxiliary supervision approaches. ``Auxiliary Supervision" = trained with extra supervision, clinical prediction, and dense fusion. The baseline late joint fusion approach is highlighted in gray.}
    \label{tab:results}
\end{table}

\subsection{Experimental Setup}

In our first experiment, we train a late joint fusion model (Section \ref{sec:fusion_methods}) with (i) concatentation, (ii) Kronecker product fusion, (iii) concatenation + all three proposed auxiliary supervision methods, and (iv) Kronecker fusion + all three auxiliary supervision methods. A comparison of these four models is performed to reveal the tradeoff between model complexity and generalization when applied to our low-dimensional dataset. To understand the contribution of each proposed technique, we also perform an ablation-style experiment that progressively adds each technique to the baseline of late joint fusion with concatenation. With the same hyperparameters and an otherwise identical architecture, we train the baseline model plus all possible combinations of the three proposed approaches: extra supervision (Section \ref{sec:extra_supervision}), clinical prediction (Section \ref{sec:clinical_prediction}), and dense fusion (Section \ref{sec:dense_fusion}). We adopt five-fold cross-validation for model training and use area under the receiver operating characteristic curve (AUC) for evaluation. Performance is summarized by the mean and standard deviation of AUC across all five folds.

\section{Results}


Compared to the standard late joint fusion approach, our proposed auxiliary supervision methods considerably improve DM classification performance (Table 1). Our proposed model (row 3) reaches 0.792 $\pm$ 0.014 AUC, while late joint fusion achieves 0.781 $\pm$ 0.024 AUC with concatenation and 0.770 $\pm$ 0.018 AUC with Kronecker product fusion. We find that Kronecker fusion increases model capacity by 4-5$\times$ the number of parameters with a consistent adverse affect on generalization in our setting. While our proposed methods \textit{also} increase model capacity, the benefits of auxiliary supervision are demonstrated by the fact that our proposed model (row 3) outperforms late joint Kronecker fusion (row 2).

\begin{table}[!ht]
    \centering
    \renewcommand{\arraystretch}{1.1}

    \begin{NiceTabular}{@{}cccc@{}}[colortbl-like]
    \toprule
    \makecell{\textbf{Extra}\\\textbf{Supervision}} & \makecell{\textbf{Clinical}\\\textbf{Prediction}} & \textbf{Dense Fusion} & \textbf{AUC} \\ \midrule
    \rowcolor{Gray} \ \ & & & 0.781 $\pm$ 0.024\ \  \\
    \ \ \checkmark & & & 0.778 $\pm$ 0.021\ \  \\
    \ \  & \checkmark & & 0.789 $\pm$ 0.013\ \  \\
    \ \ & & \checkmark & 0.776 $\pm$ 0.025\ \  \\
    \ \ \checkmark & \checkmark & & 0.787 $\pm$ 0.015\ \  \\
    \ \  \checkmark & & \checkmark & 0.785 $\pm$ 0.016\ \  \\
    \ \   & \checkmark & \checkmark & \underline{0.790 $\pm$ 0.012}\ \  \\
    \ \ \checkmark & \checkmark & \checkmark & \textbf{0.792 $\pm$ 0.014}\ \  \\ \bottomrule
    \end{NiceTabular}



    \caption{Distant metastasis (DM) classification results with all combinations of the proposed approaches. Best results appear in bold, and second-best results are underlined. The baseline late joint fusion approach is highlighted in gray.}
    \vspace*{-2mm}
    \label{tab:ablation}
\end{table}

In our ablation study, we find that best results are achieved with all three proposed auxiliary supervision approaches (Table \ref{tab:ablation}). While the baseline reaches 0.781 $\pm$ 0.024 AUC, a model that additionally uses extra supervision, clinical prediction, and dense fusion reaches 0.792 $\pm$ 0.014 AUC. Interestingly, while the addition of \textit{only} extra supervision or \textit{only} dense fusion does not improve performance, any combination of the two approaches improves upon the baseline. We also find that clinical prediction is the single most impactful of the three techniques; the three models trained with clinical prediction achieved the three largest mean AUCs and three smallest standard deviations in AUC across folds.

\section{Discussion \& Conclusion}

In summary, we proposed three simple approaches for improved deep multimodal fusion on low-dimensional datasets. Rather than designing highly expressive fusion or attention operations, the proposed techniques are less likely to overfit because they aid optimization with auxiliary sources of supervision rather than adding significant extra learning capacity. We validate these three approaches -- extra supervision, clinical prediction, and dense fusion -- on the task of learning jointly from digital histopathology imaging and tabular clinical data to predict prostate cancer metastasis, observing that the three proposed approaches improve upon the standard approach of late joint fusion. Further, the proposed approaches outperform the recently popular Kronecker product fusion while being more parameter-efficient. Though these methods were validated on one specific application and dataset, they can be readily applied to any multimodal classification task with paired image and non-image data.

This study could be expanded to perform more extensive validation of our fusion approaches on other classification tasks and even in other domains beyond hisopathology. While the combination of all three proposed methods achieved best performance, it remains to be seen whether this is the optimal combination of methods across different tasks and datasets. Lastly, our study could potentially benefit from more thorough evaluation using metrics that are resistant to class imbalance, such as average precision or balanced accuracy. We chose AUC as a metric to match the evaluation of Esteva \textit{et al.} \cite{esteva2022prostate} and avoid the need to choose a decision threshold, though AUC can become inflated with class imbalance \cite{fernandez2018learning}. It would also be valuable to compare our approaches with other baselines such as human expert performance. Further, a feature importance analysis to uncover which clinical features are most predictive of DM in this multimodal setting could enrich the clinical value of our findings.

{\small
\section{Compliance with Ethical Standards}

IRB approval was obtained my NRG Oncology through IRB00000781. Informed consent was waived because all data was anonymized.

\section{Acknowledgments}

This project was supported by U10CA180822 (NRG Oncology SDMC), U10CA180868 (NRG Oncology Operations), and \\U24CA196067 (NRG Specimen Bank) from the National Cancer Institute.

\bibliographystyle{IEEEbib}
\bibliography{ref}
}

\end{document}